# Observation of Kondo Effect in Rhombohedral Graphene Superlattices

Xin Liao[1*], Qing Yin[1*], Huiwen Wang[2*], Jing-Wei Dong[1*], Jun-Xi Chen[1], Si-Li Wu[2], Cai-Zhen Li[2†], Guowei Lyu[1], Kenji Watanabe[3], Takashi Taniguchi[4], Wei Jiang[2†], Yu-Gui Yao[2] and Zhi-Min Liao[1,5†]

[1]State Key Laboratory for Mesoscopic Physics and Frontiers Science Center for Nano-optoelectronics, School of Physics, Peking University, Beijing 100871, China

[2]Centre for Quantum Physics, Key Laboratory of Advanced Optoelectronic Quantum Architecture and Measurement (MOE), School of Physics, Beijing Institute of Technology, Beijing 100081, China

[3]Research Center for Electronic and Optical Materials, National Institute for Material Sciences, 1-1 Namiki, Tsukuba 305-0044, Japan

[4]Research Center for Materials Nanoarchitectonics, National Institute for Material Sciences, 1-1 Namiki, Tsukuba 305-0044, Japan

[5]Hefei National Laboratory, Hefei 230088, China

* These authors contributed equally to this work.

† Corresponding authors, Email: licaizhen@bit.edu.cn; wjiang@bit.edu.cn; liaozm@pku.edu.cn

## Abstract

Kondo effect in strongly correlated systems arises from the antiferromagnetic coupling between itinerant conduction electrons and localized magnetic moments, giving rise to a variety of exotic quantum phenomena. Two-dimensional moiré superlattice systems provide a highly tunable platform featuring topological flat bands, where Wannier orbitals are spatially confined by the periodic moiré potential and serve as localized magnetic moments, enabling the observable Kondo effect. Here we experimentally demonstrate Kondo interactions in hexalayer rhombohedral graphene moiré superlattices through magneto-transport and temperature-dependent measurements. With increasing magnetic field, the magnetoresistance exhibits an increase-decrease transition across a critical field $B_c$, while the Hall resistance $R_{xy}$ undergoes a sign reversal near $B_c$. Moreover, as temperature decreases, the longitudinal resistance $R_{xx}$ first increases logarithmically and then decreases following a $T^2$ behavior, indicating a transition to heavy fermion liquid. These behaviors can be consistently explained by the breakdown of Kondo singlets induced by either magnetic field or temperature, which liberates carriers previously bound to localized moments, thereby enhancing conductivity and altering the dominant carrier type. Furthermore, our results demonstrate that the Kondo interaction can be continuously tuned by both carrier density $n$ and displacement electric field $D$, and suggest the emergence of a Kondo insulating state. Our findings provide deep insight into the Kondo effect in moiré engineered flat-band systems, paving the path for exploring exotic correlated quantum phases.

## Main text

Kondo interactions between localized and itinerant electrons give rise to a variety of intriguing phenomena, including unconventional superconductivity and (topological) Kondo insulating states [1-4]. Below a characteristic Kondo temperature, the localized $f$-electrons couple antiferromagnetically with the itinerant $c$-electrons, leading to the formation of Kondo singlets [1,5,6]. When the localized magnetic moments are periodically arranged, known as a Kondo lattice, hybridization between the $f$- and $c$-electron bands opens an energy gap [7,8]. Depending on whether the Fermi level lies inside the gap, the system exhibits either a heavy Fermi liquid state or an insulating ground state, referred to as a Kondo insulator [9,10]. Recent theories and experiments have demonstrated the emergence of heavy fermions and Kondo lattices in transition metal dichalcogenide (TMD) superlattices [11-20] and magic-angle bilayer [21-33] or trilayer graphene [34,35] systems. In two-dimensional (2D) moiré superlattices, Wannier orbitals are localized at high-symmetry stacking regions, effectively serving as the $f$-electrons in a Kondo lattice. Rhombohedral graphene (Rgr) and its moiré superlattices with hexagonal boron nitride (hBN) have recently attracted considerable attention for hosting a rich variety of correlated quantum phases, including superconductivity [36-42], integer and fractional Chern insulators [42-59], and anomalous Hall crystal [51,52]. The moiré potential in Rgr/hBN superlattices can induce topological flat bands ($f$-electrons) that hybridize with dispersive bands ($c$-electrons), offering a versatile platform for realizing a highly tunable Kondo lattice. However, experimental evidence of Kondo effect in Rgr-based systems remains elusive.

In this work, we demonstrate the effect of Kondo correlations in hexalayer Rgr/hBN moiré superlattices. At filling factor $\nu = n/n_0 = 1$ (where $n_0$ is the charge density corresponding to one electron per moiré unit cell), an integer Chern insulating state appears in the moiré-distant regime, while complex anomalous Hall effects arise in the moiré-proximal regime, where electrons are pushed away from or toward the moiré interface by the applied displacement electric field, respectively. Near $\nu = 2$ in the moiré-proximal regime, a nonmonotonic magnetoresistance and a distinct Hall anomaly emerge, consistent with signatures of Kondo breakdown [13]. The applied Zeeman field disrupts Kondo singlets and releases conduction carriers [10], resulting in a decrease in magnetoresistance and a sign reversal of Hall resistance above a critical field $B_c$. The temperature dependence of longitudinal resistance $R_{xx}$ further reveals a heavy fermi liquid state at low temperature and Kondo scattering at relatively high temperature, indicating thermal decoupling of Kondo singlets [11,34]. Additionally, the strength of Kondo interaction can be continuously tuned by both carrier density $n$ and displacement electric field $D$, suggesting the emergence of a Kondo insulating state.

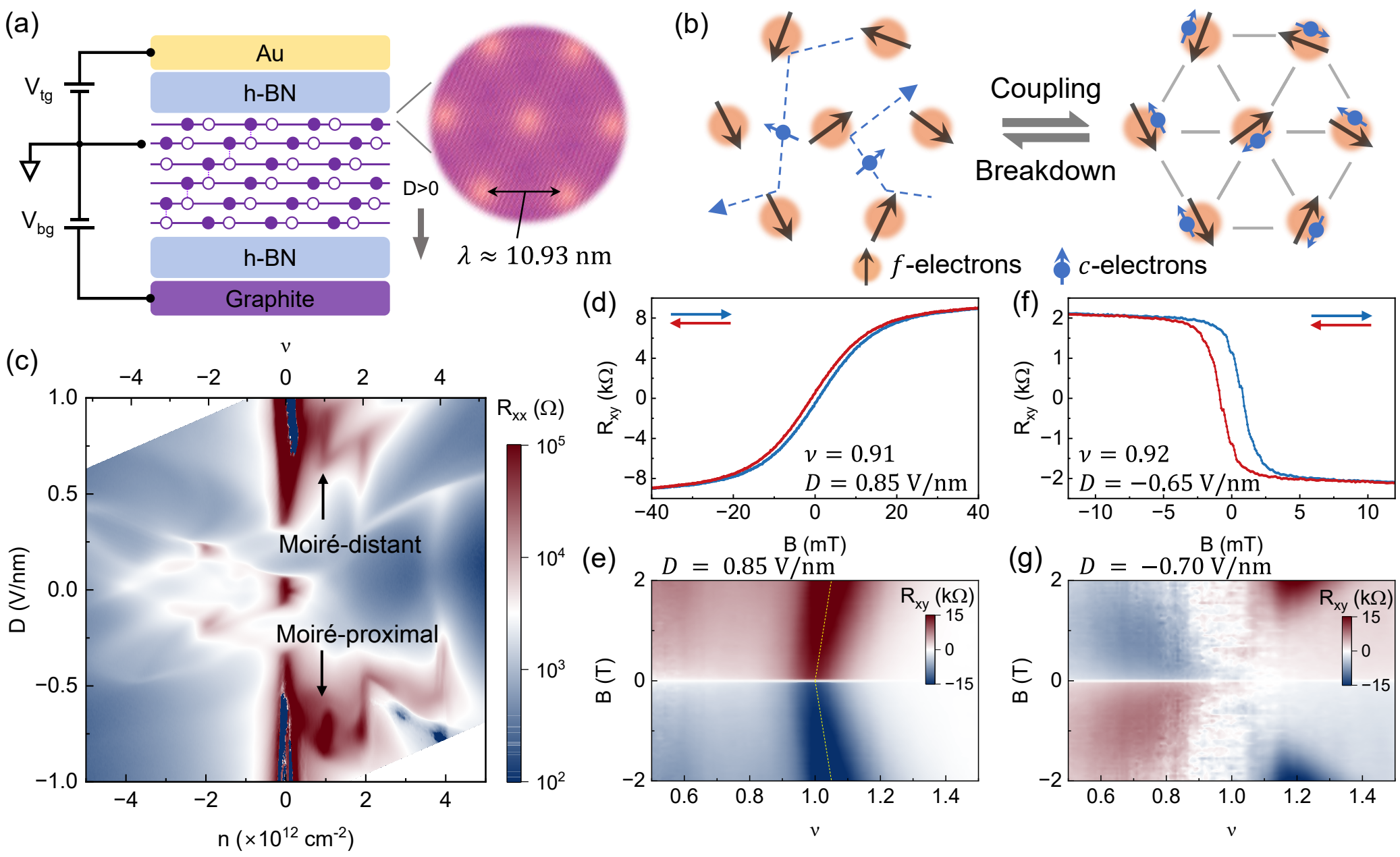

FIG. 1. (a) Schematic of the dual-gated Rgr/hBN moiré superlattice device. The hexalayer Rgr is encapsulated between two hBN flakes, forming a moiré superlattice with the upper hBN, and a moiré period of 10.93 nm. The grey arrow illustrates the direction of positive $D$. (b) Left panel: Schematic of localized moments forming a hexagonal lattice, surrounded by itinerant electrons. Right panel: After Kondo screening, the localized moments couple with itinerant electrons, forming a Kondo lattice. (c) Longitudinal resistance $R_{xx}$ map as functions of $n$ and $D$ at $B = 0$ T. (d, e) Representative magnetic hysteresis loop (d) and Landau Fan diagram of Hall

resistance $R_{xy}$ (e) measured on moiré-distant side. The yellow dashed line in (e) corresponds to the $C = 1$ state determined from the Streda formula $\partial n/\partial B = Ce/h$, where $e$ and $h$ denote electron charge and Planck constant, respectively [60]. (f, g) Same measurements as in (d, e) performed on moiré-proximal side. Data acquired at 1.5 K.

Figure 1(a) presents a schematic of the dual-gated Rgr/hBN moiré superlattice device, in which the Rgr is aligned with the top-layer hBN. The top and bottom gate voltages ($V_{tg}$ and $V_{bg}$) independently control $n$ and $D$ (see Supplementary Note 1 [61]). The lattice mismatch, along with a small twist angle $\theta$ between Rgr and hBN layers, generates moiré patterns that forms triangular lattice. The strong periodic moiré potential gives rise to flat band, which overlaps with the higher Dirac-like dispersion band on the electron-doped side at $D = 0\ \mathrm{V/nm}$ (see Supplementary Note 2 [61]). Upon applying a fixed $D$, the lowest conduction band progressively flattens, and the hybridization-driven reconstruction between this flat band and the Dirac band leads to the opening of a gap, resulting in the itinerant electrons interact with localized moments via Kondo hybridization and the formation of a correlated Kondo phase [Fig. 1(b)].

Figure 1(c) shows the longitudinal resistance $R_{xx}$ map as functions of $n$ and $D$ at a temperature of $T = 1.5$ K and zero magnetic field. The twist angle is estimated to be $\theta = 0.87^{\circ}$ (corresponding to a moiré periodicity $\lambda \approx 10.93$ nm), from the full filling carrier density $n_s = 4n_0 \approx 3.87 \times 10^{12}\ \mathrm{cm}^{-2}$ (see Supplementary Note 1 [61]). The relatively large twist angle leads to a weaker moiré potential compared with previously reported Rgr/hBN moiré superlattice [55-57]. As a result, the quantum phases at the charge neutrality point remain similar to those in moiré-free Rgr, including the layer antiferromagnetic and layer-polarized insulating states [43,71]. On the electron-doped side, three primary resistance peaks appear under both positive and negative $D$. The peak at $\nu = 4$ corresponds to band insulators resulting from full filling of the first conduction miniband, while the peaks at $\nu = 1$ and $2$ arise from correlations [72] (see Supplementary Fig. S5 [61]).

As shown in Fig. 1(d-g), Chern insulating states near $\nu = 1$ are observed in both the moiré-distant and moiré-proximal regimes, and extend over a broad range of $\nu$ and $D$ (see Supplementary Fig. S6 [61]). In the moiré-distant regime, the Chern insulating state with Chern number $C = 1$ is confirmed by the Landau fan diagram, where the maxima of Hall resistance $R_{xy}$ follows the Streda formula [60]. By contrast, the moiré-proximal regime exhibits opposite magnetic hysteresis loop that does not conform to the Streda formula, suggesting strong electronic correlations and complex competing mechanisms in this regime [58]. Overall, the magnetic hysteresis loop and anomalous Hall effect, consistent with prior experimental results [42,49,51-58], indicate spontaneous time reversal symmetry breaking and intrinsic orbital magnetism [73], suggesting the high quality of our device.

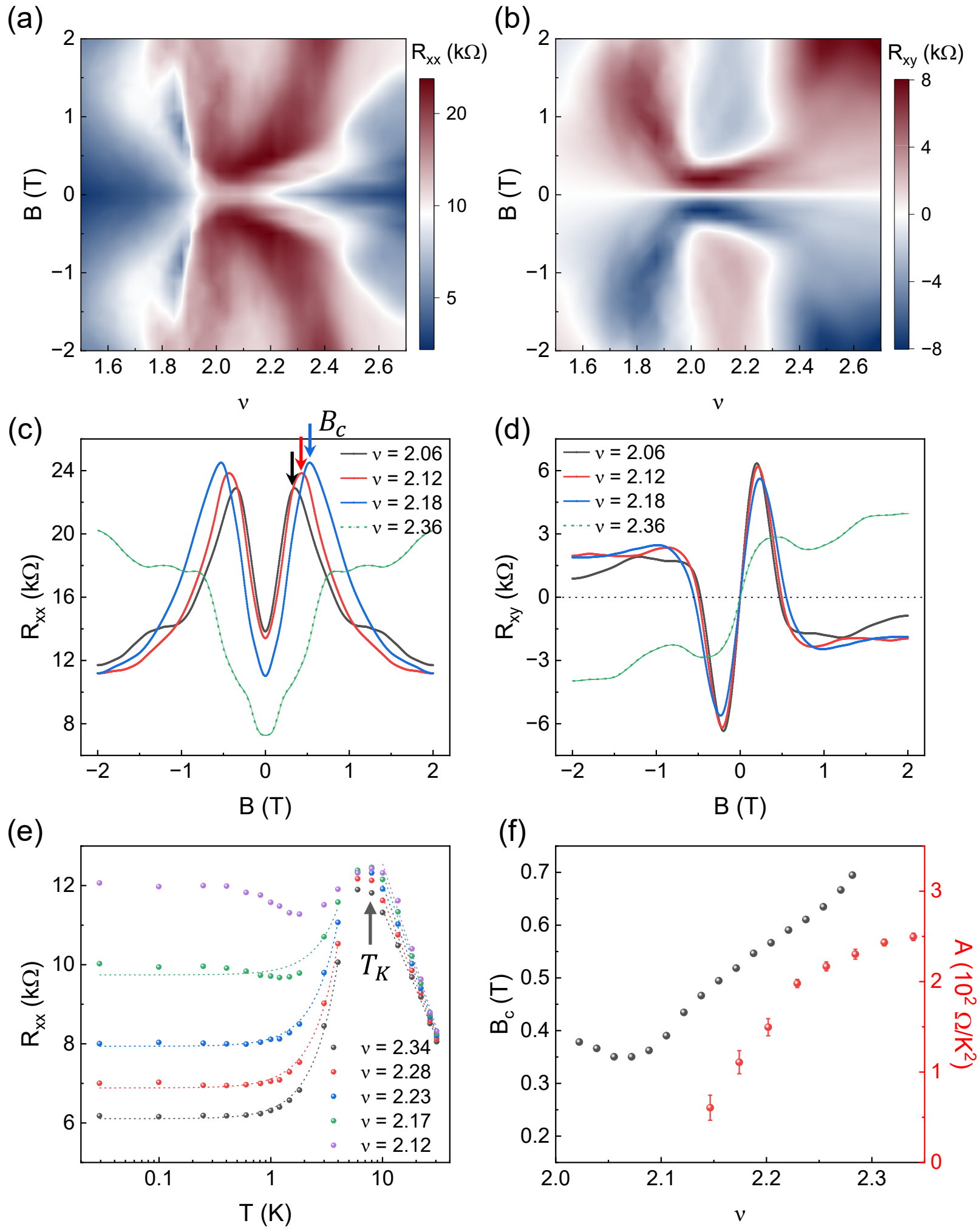

FIG. 2. (a, b) Symmetrized $R_{xx}$ (a) and anti-symmetrized $R_{xy}$ (b) maps as functions of $\nu$ and $B$ at $D = -0.4275$ V/nm and $T = 30$ mK. Nonmonotonic magnetoresistance and Hall anomaly emerge within the range of $2 < \nu < 2.3$. (c, d) Line cuts of $R_{xx}$ (c) and $R_{xy}$ (d) versus $B$ at representative filling factors extracted from (a) and (b), respectively. Arrows in (c) indicate the critical field $B_c$ for each curve. (e) Temperature dependence of $R_{xx}$ at various filling factors and $B = 0$. Arrow marks the Kondo temperatures $T_K$. Dashed lines below $T_K$ represent fits to the quadratic temperature dependence, while those above $T_K$ correspond fits to logarithmic temperature dependence. (f) Extracted $B_c$ and coefficient $A$ as a function of $\nu$.

We now focus on the correlated state near $\nu = 2$ in the moiré-proximal regime. Figure 2(a-b) present $R_{xx}$ and $R_{xy}$ maps as functions of $\nu$ and perpendicular magnetic field $B$ at $D = -0.4275$ V/nm and $T = 30$ mK. Pronounced nonmonotonic magnetoresistance and Hall anomaly emerge within the range of $2 < \nu < 2.3$. Line cuts of $R_{xx}$ and $R_{xy}$ versus $B$ at representative filling factors are illustrated in Fig. 2(c-d). With increasing $B$, $R_{xx}$ first rises rapidly, reaches a maximum near a critical field $B_c$, and subsequently decreases, forming an M-shaped curve. Simultaneously, $R_{xy}$ exhibits a similar evolution and undergoes a sign reversal near $B_c$. In contrast, away from this regime, the resistance increases monotonically with $B$ and no sign reversal is observed in $R_{xy}$. These behaviors are reminiscent of those observed in Kondo lattice systems [11,13]. In general, Kondo singlets form in heavy fermion systems below the characteristic Kondo temperature [5]. An increasing Zeeman field destroys the antiferromagnetic coupling between the itinerant electron spins and the localized moments (including spin, valley and orbital), which is a process known as Kondo breakdown [13]. This disruption releases itinerant carriers, resulting in a reduction of $R_{xx}$. The subsequent change in the dominant carrier type from electrons to holes gives rise to the sign reversal of $R_{xy}$, indicating the emergence of hole pockets due to magnetic field-induced Fermi surface reconstruction (see Supplementary Fig. S8 [61]). Beyond this transition, quantum oscillations appear immediately due to the sudden increase in carrier density (see Supplementary Fig. S7 [61]).

To further reveal the Kondo effect in our system, we examine the temperature dependence of $R_{xx}$ at various $\nu$ under zero magnetic field ($D = -0.4275$ V/nm), as shown in Fig. 2(e). As temperature decreases, a pronounced insulator-to-metal transition occurs at a critical temperature $T_K \sim 8$ K, the Kondo temperature, signaling the onset of Kondo screening. Above $T_K$, $R_{xx}$ exhibits $-\log(T)$ dependence, which persists up to 180 K, significantly different from the non-Kondo region (see Supplementary Figs. S9 and S11 [61]). This behavior can be attributed to spin-flip scattering mediated by Kondo effect [74]. Below $T_K$, the localized moments become progressively screened by conduction electrons, resulting in a metallic transition and Fermi liquid behavior that can be described as $R_{xx} = R_0 + AT^2$, where $R_0$ is the residual resistance and $A$ is the Kadowaki-Woods coefficient, proportional to the square of quasiparticle mass $m^*$ [$A \propto (m^*)^2$] [75]. The extracted coefficient $A$ (up to $\sim 250\ \Omega/\mathrm{K}^2$), shown in Fig. 2(f), is an orders of magnitude larger than that obtained from other pure Fermi liquid regions of our system (see Supplementary Note 4 [61]), indicating a substantial enhancement of quasiparticle effective mass.

Figure 2(f) shows the dependence of $B_c$ and the coefficient $A$ on $\nu$. The decrease of $B_c$ with decreasing $\nu$ results from the reduced screening effect of itinerant electrons on local magnetic moments, which weakens the Kondo interaction [11]. Meanwhile, as $\nu$ approaches 2, $R_{xx}$ deviates from the $T^2$ dependence and reenters an insulating state [Fig. 2(e)], leading to the inevitable decrease of coefficient $A$ and even failure of fitting. Similar behavior has been reported in $MoTe_2/WSe_2$ moiré bilayers due to the formation of topological insulating states [13]. Therefore, near $\nu = 2$, the effective mass and the strength of Kondo coupling cannot be fully captured by the coefficient $A$ alone. Additional interaction mechanisms are likely required to account for the insulating behavior.

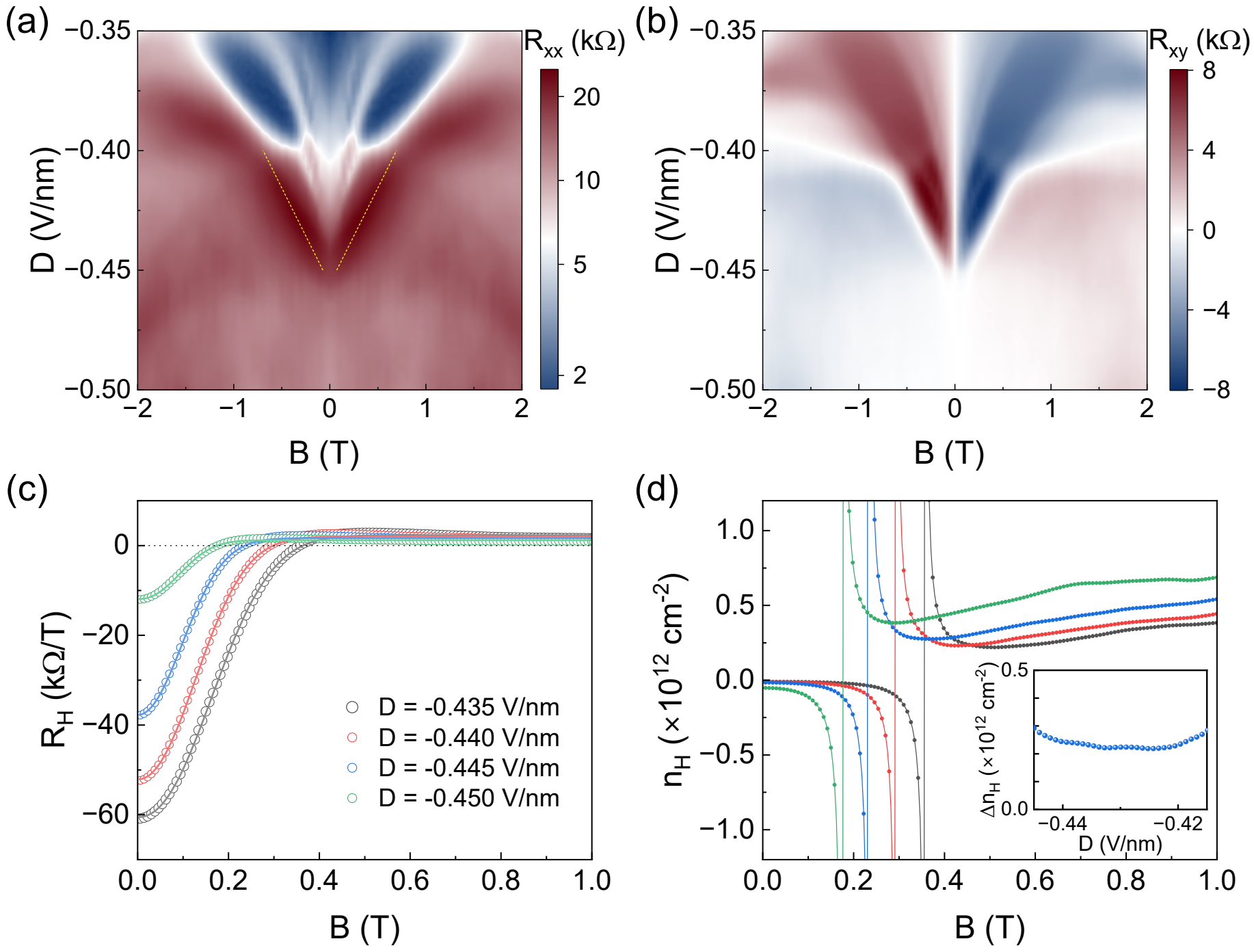

FIG. 3. (a, b) Symmetrized $R_{xx}$ (a) and anti-symmetrized $R_{xy}$ (b) maps as functions of $D$ and $B$ at $\nu = 2.07$ and $T = 250$ mK. The dashed yellow lines in (a) mark the linear fit of the critical field $B_c$ versus $D$, consistent with the Hall sign reversal observed in (b). (c, d) Hall coefficient $R_H$ (c) and Hall carrier density $n_H$ (d) as functions of $B$, extracted from line cuts at selected $D$ in (b). The circles in (c) represent $R_H$ calculated by $R_H = R_{xy}/B$, while the solid curves denote fits to $R_H = R_H^0 + A^* \exp(-\eta B^\beta)$. Inset in (d) shows $\Delta n_H$, i.e. the variation of Hall carrier density across the abrupt jump at $B_c$, under varying $D$.

Next, we demonstrate a continuously $D$-tunable Kondo effect. Figure 3(a-b) display maps of $R_{xx}$ and $R_{xy}$ versus $D$ and $B$ at $\nu = 2.07$ and $T = 250\ \mathrm{mK}$. Three regimes are distinguished here (see also Supplementary Fig. S10 [61]). For $D > -0.4\ \mathrm{V/nm}$, quantum oscillations appear at low magnetic field, where the dips in $R_{xx}$ coincide with the quantum Hall plateaus of $R_{xy}$ and no sign reversal is observed in $R_{xy}$. For $-0.4\ \mathrm{V/nm} > D > -0.45\ \mathrm{V/nm}$, the critical field $B_c$ decreases almost linearly with increasing electric field $|D|$ and vanishes near $D \approx -0.45\ \mathrm{V/nm}$, as indicated by the guide line in Fig. 3(a). For $D < -0.45\ \mathrm{V/nm}$, the further increased electric field tunes the energy offset between layers, effectively reducing the Kondo interaction and forming a Mott insulator near $\nu = 2$, due to the increasing ratio of on-site Coulomb repulsion to bandwidth ($U/W$) [75] (see Supplementary Note 2 [61]). Meanwhile, in the moiré-distant regime, no Hall anomaly is detected (see Supplementary Fig. S6 [61]), since weak moiré potential does not support the formation of localized moments.

Hall anomaly has been previously reported in various contexts, but conventional mechanisms fail to explain our data. First, while coexisting electrons and holes can cause nonlinear magnetoresistance and Hall sign reversal in semimetals (e.g., $WTe_2$ [76] and $Cd_3As_2$ [77]), multi-band models cannot fit our results. Second, alternative symmetry-broken states near half-filling are inconsistent with our observations: the absence of magnetic hysteresis and a Chern insulating state rules out a valley-polarized ferromagnetic state; a conventional spin-polarized half-metal fails to capture the distinct field-induced transition; and reported intervalley-coherent states [57] exhibit entirely different magnetotransport behaviors. Finally, Hartree-Fock calculations [58] favor a spin-polarized, valley-balanced configuration, rendering opposite-spin antiferromagnetic pairing unlikely. Therefore, standard symmetry-broken scenarios cannot fully account for our findings (see Supplementary Note 6 [61] for detailed discussions).

Furthermore, the Kondo singlets can be destroyed by in-plane magnetic field with a critical field $B_{\parallel c}$, nearly ten times larger than the out-of-plane critical field $B_{\perp c}$, indicating the role of orbital degree of freedom (see Supplementary Fig. S15 [61]). Given the large valley (orbital) $g_v$ factor and the relatively small spin $g_s$ [56], the Zeeman energy scale $g_v \mu_B B_{\perp c}$ or $g_s \mu_B B_{\parallel c}$ ($\mu_B$ is the Bohr magneton) becomes comparable to the Kondo temperature $k_B T_K$ ($k_B$ is the Boltzmann constant).

To gain more insight into the Hall anomaly, an empirical function is used to fit the Hall coefficient [78]:

$$R_H = R_H^0 + A^* \exp(-\eta B^\beta),$$

as shown in Fig. 3(c), where $R_H^0$ is a field-independent term and $A^* \exp(-\eta B^\beta)$ is the anomalous contribution term. Significantly, a similar relationship has been observed in Kondo insulator $YbB_{12}$ [78] and iron-based superconductors [79,80], associated with proximity to the quantum critical point. Thus, the successful fit with $\beta \sim 2$ strongly indicates non-Drude physics within the Kondo-screened phase. As shown in Fig. 3(d), Hall carrier density, calculated from the Hall coefficient via $n_H = 1/(eR_H)$, exhibits an abrupt jump over $B_c$, consistent with the breakdown of the Kondo singlet. The variation of Hall density remains nearly constant, $\Delta n_H \sim 2.5 \times 10^{11}\ \mathrm{cm}^{-2}$, over varying $D$ at fixed filling factor [inset of Fig. 3(d)]. Subsequently, the released itinerant carriers take part in quantum oscillations.

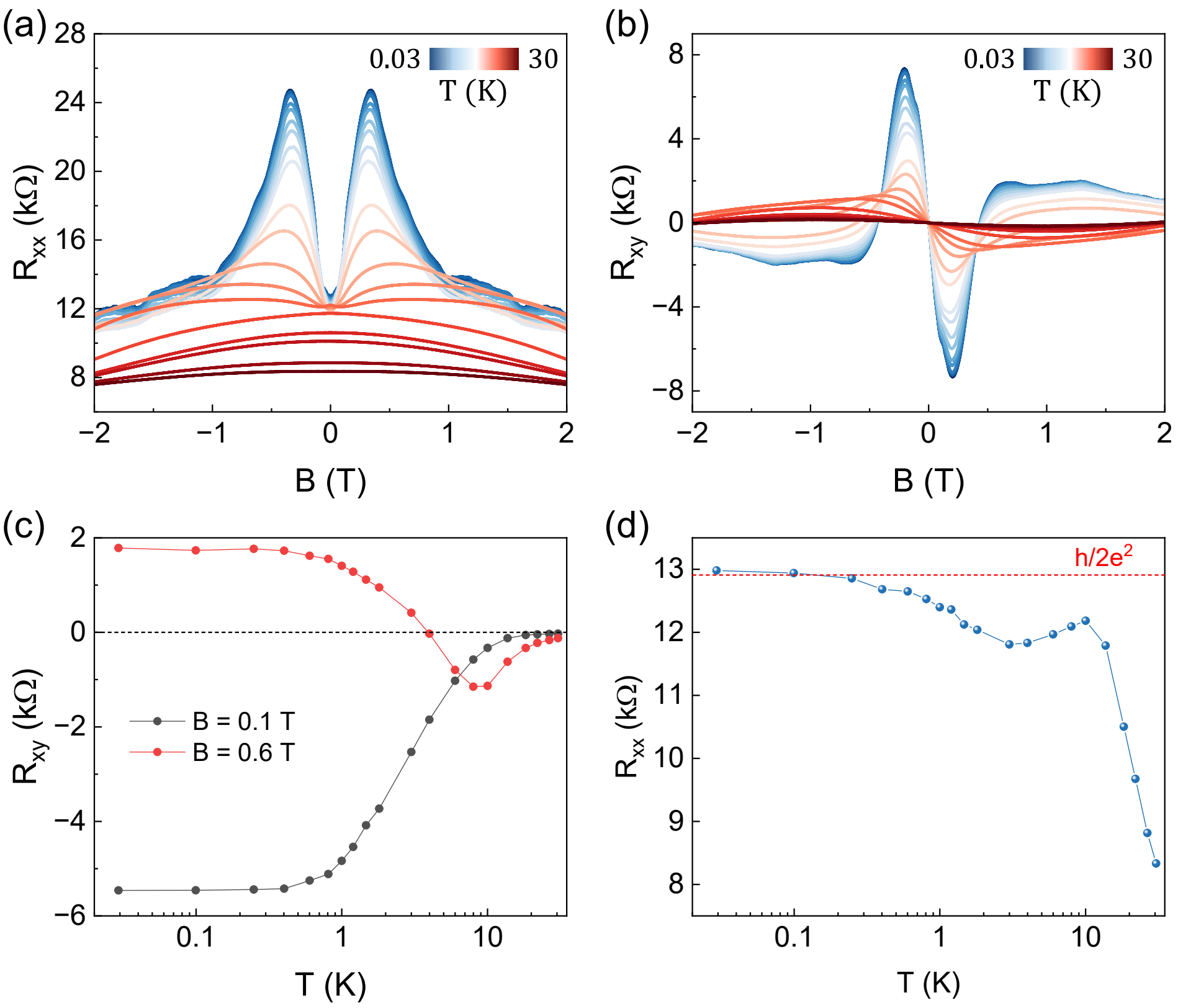

FIG. 4. (a, b) Symmetrized $R_{xx}$ (a) and anti-symmetrized $R_{xy}$ (b) as functions of $B$ at various temperatures for $\nu = 2.07$ and $D = -0.4275\ \mathrm{V/nm}$. (c) $R_{xy}$ at $B = 0.1\ \mathrm{T}$ and $0.6\ \mathrm{T}$ as a function of $T$, extracted from (b). (d) $R_{xx}$ at $B = 0$ as a function of $T$, extracted from (a). As shown, $R_{xx}$ saturates to a quantized plateau of $h/2e^2$ in the low temperature limit.

Finally, we examine the $B$ dependence of $R_{xx}$ and $R_{xy}$ at various temperatures for $\nu = 2.07$ and $D = -0.4275\ \mathrm{V/nm}$ [Fig. 4(a-b)]. As the temperature decreases, the M-shaped magnetoresistance profiles in $R_{xx} - B$ curves first emerge below $T_K$, consistent with the first phase transition observed in $R_{xx} - T$ curve [Figs. 2(e) and 4(d)], which can be attributed to the onset of screening of localized moments by itinerant electrons. Upon further cooling, Hall anomaly with sign reversal appears near 4 K, suggesting the occurrence of a meta-magnetic phase transition, similar to those reported in rare-earth heavy fermion compounds [81] and Kondo lattice systems in $MoTe_2/WSe_2$ moiré bilayers [11]. Moreover, the extracted $R_{xy}$ at $B = 0.1$ T (before Kondo breakdown) decreases monotonically with temperature and saturates at low temperature [Fig. 4(c)], implying a contribution from localized moments and ruling out extrinsic skew scattering [34,82]. In contrast, $R_{xy}$ at $B = 0.6$ T, above the critical magnetic field $B_c \sim 0.35$ T, reaches a minimum near 8 K, followed by an increase and sign reversal around 4 K, which can be ascribed to asymmetric scattering of decoupled conduction carriers by the localized moments. Notably, the local $R_{xx}$ saturates to a quantized plateau of $h/2e^2$ in the low temperature limit [Fig. 4(d)], suggesting the possible emergence of a topological insulator with helical edge states [13]. While theoretical works have proposed a quantum spin Hall state at $\nu = 2$ [83,84], the extremely small energy gap restricts the observation of this state and additional experimental evidence is required to confirm this scenario (see Supplementary Note 5 [61]).

In conclusion, we have presented strong evidence for Kondo interactions in Rgr/hBN moiré superlattices. Near $\nu = 2$ in the moiré-proximal regime, Kondo breakdown induced by the Zeeman field emerges, leading to a sharp drop in magnetoresistance and a pronounced Hall anomaly. Moreover, $R_{xx}$ exhibits heavy fermion liquid behavior at low temperature and logarithmic negative temperature dependence at high temperature, indicating the emergence of heavy fermions and thermal induced decoupling of Kondo singlets. These behaviors evolve systematically with carrier density, displacement electric field, and temperature. Although a direct characterization of the low-field Fermi surface by quantum oscillations remains experimentally challenging because of the strongly enhanced quasiparticle mass, the combined transport observations are most consistent with the Kondo scenario. Our work enriches the quantum phase landscape of moiré flat-band systems, opening new avenues for exploring strongly correlated and topological phenomena.

## Acknowledgements

This work was supported by National Natural Science Foundation of China (Grant Nos. 62425401, 12534001, 12504044, 62321004 and 12321004), and Quantum Science and Technology-National Science and Technology Major Project (Grant No. 2021ZD0302403). K.W. and T.T. acknowledge support from the JSPS KAKENHI (Grants No. 20H00354, No. 21H05233, and No. 23H02052) and World Premier International Research Center Initiative (WPI), MEXT, Japan.